\documentclass{aa}
\usepackage{txfonts}
\usepackage{graphicx}
\usepackage{natbib}
\usepackage{longtable,lscape}
\bibpunct{(}{)}{;}{a}{}{,}
\begin{document}
\title{Wavelet analysis of a large sample of AGN at high radio frequencies}

\author{T. Hovatta \inst{1} \and H.J. Lehto \inst{2,3} \and M. Tornikoski \inst{1}}
\institute{Mets\"ahovi Radio Observatory, TKK, Helsinki University of Technology, Mets\"ahovintie 114, 02540 Kylm\"al\"a, Finland  \\ \email{tho@kurp.hut.fi} \and Tuorla Observatory, University of Turku, V\"ais\"al\"antie 20, 21500 Piikki\"o, Finland \and Department of Physics, 20140 University of Turku, Finland}
\date{Received / Accepted}
\abstract
{}
{We have studied the characteristic timescales of 80 AGNs at 22, 37 and 90\,GHz examining the properties of the wavelet method and comparing them to traditional Fourier-based methods commonly used in astronomy.}
{We used the continuous wavelet transform with the Morlet wavelet to study the characteristic timescales. We also gain information when the timescale is present in the flux curve and if it is persistent or not.}
{Our results show that the sources are not periodic and changes in the timescales over a long time are common. The property of wavelets to be able to distinguish when the timescale has been present is superior to the Fourier-based methods. Therefore we consider it appropriate to use wavelets when the quasi-periodicities in AGNs are studied.}
{}

\maketitle
\section{Introduction}
Wavelet methods have been used for the past 10 years to study 
the variability timescales in different types of astronomical 
objects \citep[e.g.][]{foster96, priestley97, scargle97}. 
Wavelets have properties superior to 
Fourier based methods, the most useful being the locality of the 
timescale.
In Fourier based methods a single sinusoid is fitted to the whole light curve, whereas wavelets enable us to study temporally local timescales or periods that are present in the light curves. 
This is a very good property when studying flux curves of AGNs, which 
are almost never strictly periodic over long times. 

Wavelets were used to study variability timescales of a larger 
sample of AGNs by \cite{kelly03}. They used the cross-wavelet transform 
to study quasi-periodic variations in the Pearson-Readhead VLBI 
sample, monitored by the University of Michigan Radio Observatory (UMRAO). 
A quasi-periodic behaviour was found in a little over half of the 30 sources 
they studied. Individual sources have also been studied for quasi-periodicity 
and periodicity using wavelets \citep[e.g.][]{hughes98, lehto99, kadler06, ciprini07}.

In \cite{hovatta07}, hereafter Paper I, we studied the variability 
timescales of a large sample of AGNs using the structure function (SF), 
the discrete autocorrelation function (DCF) and the Lomb-Scargle 
periodogram. We found 
that larger flares in these sources happen quite rarely, only once in 4--6 
years while short-term variations are continuously present in the flux 
curves. Also, many of the sources had changed their behaviour during the 
25 years of monitoring which is one reason why we wanted to study the sample 
again using wavelets. In Paper I we also studied the 
properties of the methods and their 
differences by searching for the characteristic timescales. We noticed that 
there are several problems when using periodograms to study the timescales because none of the sources in our sample showed strict periodicities 
in their flux curves or were well characterised by sinusoids. 

In this paper we will compare the results from this wavelet analysis to 
Paper I. This paper is organised as follows: In Sect. 
\ref{sect:sample} we describe the source sample and data used. Section 
\ref{sect:method} includes the description of the method and 
in Sect. \ref{sect:results} we present the results. The discussion is in 
Sect. \ref{sect:discussion} and the conclusions are drawn in 
Sect. \ref{sect:conclusions}. 

\section{The sample and observations} \label{sect:sample}
We used the same sample of 80 AGNs selected from the Mets\"ahovi monitoring 
list that was used in Paper I. The 
sample consists of 24 BL Lacertae Objects (BLOs), 23 Highly Polarised 
Quasars (HPQs), 28 Low Polarisation Quasars (LPQs) and 5 Radio Galaxies 
(GALs). 
The five sources with no information about their optical 
polarisation are considered to be LPQs in this study and are included in the 
numbers above.

All sources have been monitored for at least 10 years at 22 or 37\,GHz. 
These are all bright sources with a flux density of at least 1 Jy in 
the active state in at least one of the two frequencies. 
We used the 22, 37 and 87\,GHz data obtained 
with the Mets\"ahovi 14-metre radio telescope over 25 years of
monitoring \citep{salonen87, terasranta92, terasranta98, terasranta04, terasranta05, nieppola07}. 
The 37\,GHz data of HPQs, LPQs and GALs from December 2001 until March 
2005 are partially unpublished. The details of the observation method 
and data reduction processes are given in \cite{terasranta98}.
In addition we used 90\,GHz data obtained with the 
Swedish-ESO Submillimetre Telescope (SEST) at La Silla, Chile, which was used in our monitoring campaign to sample the high frequency, 90 and 230\,GHz variability of southern and equatorial sources
\citep{tornikoski96}.
The monitoring campaign at 
SEST lasted from 1987 to 2003 from which the data from 1994.5-2003 are unpublished. 
Complementary high frequency data at 90\,GHz were also collected from the literature \citep{Steppe88, Steppe92, Steppe93, Reuter97}. 
The median intervals between the observations at 22, 37 and 90\,GHz are 31, 41 and 47 days, respectively. The values depend on the source and at 37\,GHz the minimum average sampling rate was 6.8 days for the source \object{3C 84} which is used as a secondary calibrator in the Mets\"ahovi observations. The maximum average value at 37\,GHz was 186.4 days for the source \object{2234+282}.

\section{The wavelet method} \label{sect:method}
We have analysed the light curves using Morlet wavelets. They can be understood as local wave packets. The method is closely related to chirp analysis methods, Short Time Fourier Transform methods, Gabor transforms and also to the sonogram or spectrogram analysis used widely in audio signal processing. In wavelet analysis the length of the analysis ``package'', the kernel, is proportional to the timescale of interest, or in terms of frequency, proportional to the inverse of the frequency. We have used a Morlet wavelet, which can be understood mathematically as a localised plane wave tapered by a Gaussian function.  For longer timescales of interest we thus use lower frequency sinusoids than for higher frequencies according to the following equations. Wavelet transforms are functions of both frequency, $f$, (or the respective timescale) and the local time, $\tau .$

\begin{equation}
g^*(f,\tau)=\exp \left(-icf(t-\tau)-{1\over 2}(f(t-\tau))^2\right),
\end{equation}
where we adopt $c=2\pi ,$ which characterises the amount of tapering. The power of the wavelet transform, known as the scalogram, is defined for evenly spaced data as
\begin{equation}
W(f,\tau)=f\cdot\left( S^2(f, \tau ) + C^2(f, \tau)\right),
\end{equation}
where
\begin{equation}
S(f, \tau)=\sum_{j=1}^N m_j\sin \left(2\pi f(t_j-\tau)\right)\exp\left({-1\over 2}(f(t_j-\tau))^2\right) ,
\end{equation}
\begin{equation}
C(f, \tau)=\sum_{j=1}^N m_j\cos \left(2\pi f(t_j-\tau)\right)\exp\left({-1\over 2}(f(t_j-\tau))^2\right) , 
\end{equation}
where $m_j$ and $t_j$ are respectively the flux density and the time of the observation $j$.

For unevenly spaced data we have to weight each point with the inverse of the local density of points. 
The locality, a property of wavelets and other similar methods mentioned above, is in strong contrast to global methods such as Fourier transforms or structure functions, which treat the whole light curve as one entity. A change in a few critical points can change the final result in global methods rather significantly, whereas in wavelet analysis and other local methods a chance point in the data will only affect the analysis locally. As we gain information about the local nature of the signal we lose in the resolution with which we can determine the timescale (or frequency) of the signal. The length of the kernel used limits our resolution to about 0.07
in log space, which is significantly worse than what would be obtained e.g. from the Fourier analysis of a long lasting sinusoid with a constant period. Wavelet transforms with longer kernels could also be applied to the data as intermediate forms approaching Fourier transforms in accuracy, but in doing so one loses the locality information according to 
\begin{equation}
\Delta t \Delta \nu >\delta,
\end{equation}
which is analogous to the Heisenberg uncertainty principle. Here
$\delta$ can be considered as unity if we are interested in separating two nearby peaks. If we rather wish to measure the location of a single signal in weakly noisy data then $\delta$ is of the order of $\sqrt{(1/N)},$ where $N$ is the local number of points.

First we calculate the scalogram for our unevenly spaced data.
We then calculate a scalogram of normally distributed Gaussian white noise with a variance of unity sampled as was the original data. By repeating this 20 times we get a good estimate of the scalogram of this white noise. We then normalise our original scalogram by the white noise scalogram. If our data were pure noise, this noise-normalised scalogram would have everywhere an expected value equal to the variance of our measurements. Alternatively, we argue that by taking the square root of the noise-normalised scalogram and dividing it by the (known) measurement errors we get an estimate of the significance of the variations detected. If a variability timescale has a variance of $>9$ as expected from measurement errors, we consider that a significant detection and if the variance is between 4 and 9 times the variance of measurements, we consider that a marginal case (marked as a weak timescale in Table \ref{table:timescales}). These numbers were selected to represent ``2 and 3 sigma'' levels, but we caution that these should be considered indicative only, as we have not yet carried out full simulations of the probability density distributions of the noise properties in noise-normalised scalograms.

\section{Wavelet timescales} \label{sect:results}
\subsection{Interpretation}
The advantage of wavelets is that in addition to finding the variability timescale in the frequency domain, we can see at what moments it has been present. This enables us to see whether the timescale is present during the whole observed time or is transient. In addition we can identify typical flare timescales i.e. flare duration or rise and decay times and notice whether one large outburst dominates the timescale obtained. Also short-term variability is easily detected.

\begin{figure}
\resizebox{\hsize}{!}{\includegraphics{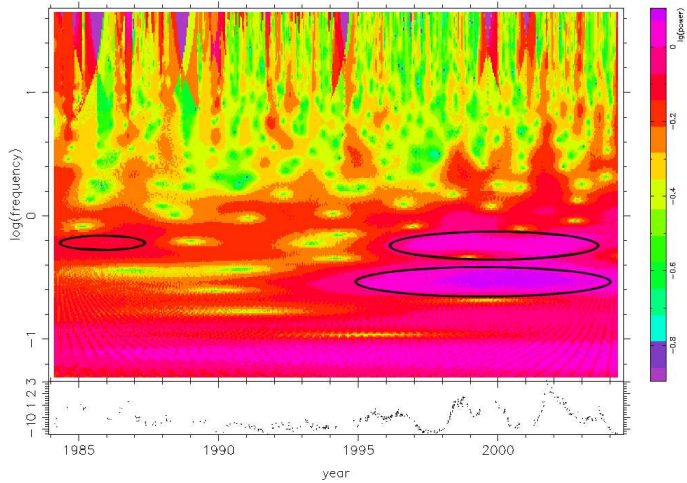}}
\caption{Wavelet transform of the source 1156+295 at 22\,GHz. The y-axis is frequency 1/year so that a timescale of 1 year corresponds to frequency 0 and a timescale of 10 years corresponds to frequency $-1$ in the log-axis. A long-term variability timescale of 3.4 years ($10^{0.53}$ years) is visible in the latter half of the flux curve starting at 1995. A flare timescale of 1.7 years ($10^{0.23}$ years) is also seen.}
\label{example}
\end{figure}

We searched for both long-term variability, which is present most of the 
time in the flux curves, and typical timescales in which the flares occur.
For the long-term timescales we first identified the timescale from 
the frequency axis and then determined how long it has been present on 
the time axis. This way we can study how many times the timescale has 
repeated during the time it has been present and also how long time it has 
been present compared to the total observing period. Figure \ref{example} 
shows the wavelet transform for the source \object{1156+295} 
(\object{4C 29.45}). We find a timescale of 3.4 years ($10^{0.53}$ years) 
which is present in the latter half of the flux curve, starting at 
1995. The timescale is marked with a 
circle in the plot.
Furthermore we 
identified typical flare timescales in the cases where more than one flare 
occurs on approximately the same timescale. For \object{1156+295} it is 
seen at 1.7 years ($10^{0.23}$ years). This timescale is also shown 
in the plot as circles at the beginning of the time series at the time of the 
first two flares and later simultaneously with the long-term trend. The 
timescale can be identified with the durations of these flares which are 
of about 1--2 years.

Often the long-term timescales are either slowly shortening or increasing in 
time which is seen as rising or declining trends in the wavelet plot. 
These are often connected to flares and usually the timescales shorten 
towards the peak of a large flare. 

\subsection{Long-term trends}
We were able to determine a long-term timescale for 122 cases out of the 
total 192 cases for which we calculated the wavelet transform. In 
addition to the timescale, we determined how long it is present 
in the flux curve. This way we were able to calculate how many times 
the cycle has repeated itself. The timescales are listed in Table 
\ref{table:timescales}\addtocounter{table}{1}, where the B1950-name, 
a well-known alias (when applicable), classification, 
total monitoring time, timescale and number of cycles and 
timescale of flares are tabulated for each frequency band. If there is more 
than one timescale present the most significant one is given first.
In 18 sources ($\sim$1/4 of the sample) we found long-lasting cycles, continuing for at least four cycles. Five of these sources showed cycles at two separate frequencies and three sources at all the three frequency bands.  
There were 31 cases where the variability timescale did not last for two cycles.
In Table \ref{table:averages} we have 
calculated averages and medians for the timescales continuing for at least two cycles.
They are given for each frequency and source class separately. In addition 
we corrected the timescales for redshift and these averages are 
also shown in Table \ref{table:averages}.

\begin{table*}
\caption[]{Averages and medians of long term timescales that have lasted for at least two cycles. 
Values are given for each frequency and source class separately. Also the redshift corrected averages are shown.}
\label{table:averages}
\centering
\begin{tabular}{llllllllll}
\hline
\hline
\noalign{\smallskip}
Freq 	  & 	 type 	  & 	 ALL 	  & 	 N 	  & 	 BLO 	  & 	 N 	  & 	 HPQ 	  & 	 N 	  & 	 LPQ 	  & 	 N 	  \\
 \noalign{\smallskip} 
\hline 
\noalign{\smallskip} 
22 	  & 	 average 	  & 	 4.6 	  & 	 37 	  & 	 5.1 	  & 	 13 	  & 	 4.8 	  & 	 10 	  & 	 4.2	  & 	 10 	  \\
  	  & 	 redshift corr. 	  & 	 3.0 	  & 	 37 	  & 	 3.6 	  & 	 13 	  & 	 2.7 	  & 	 10 	  & 	 2.4 	  & 	 10 	  \\
 	  & 	 median 	  & 	 4.3 	  & 	 37 	  & 	 5.5 	  & 	 13 	  & 	 4.9 	  & 	 10 	  & 	 3.9 	  & 	 10
 	  \\
\hline
 37 	  & 	 average 	  & 	 4.4 	  & 	 41 	  & 	 5.2 	  & 	 10 	  & 	 4.5 	  & 	 15 	  & 	 3.8 	  & 	 12 	  \\
  	  & 	 redshift corr. 	  & 	 2.8 	  & 	 40 	  & 	 4.2 	  & 	 9 	  & 	 2.4 	  & 	 15 	  & 	 2.1 	  & 	 12 	  \\
 	  & 	 median 	  & 	 4.3 	  & 	 41 	  & 	 4.5 	  & 	 10 	  & 	 4.3 	  & 	 15 	  & 	 3.7 	  & 	 12
 	  \\
\hline
 90 	  & 	 average 	  & 	 2.9 	  & 	 11 	  & 	 2.6 	  & 	 4 	  & 	 3.0 	  & 	 4 	  & 	 3.5 	  & 	 2 	  \\
  	  & 	 redshift corr. 	  & 	 1.9 	  & 	 11 	  & 	 1.6 	  & 	 4 	  & 	 1.9 	  & 	 4 	  & 	 2.3	  & 	 2 	  \\
 	  & 	 median 	  & 	 2.7 	  & 	 11 	  & 	 2.3 	  & 	 4 	  & 	 3.1 	  & 	 4 	  & 	 3.5 	  & 	 2
 	  \\
 \noalign{\smallskip} 
\hline 
\end{tabular}
\end{table*}

We ran a Kruskal-Wallis analysis to study whether the timescales at different 
frequencies and source classes differ from each other. The timescales at 
90\,GHz differ from lower frequencies with a 95\% confidence limit with 
the timescales being shorter. The result can be affected by the smaller 
number of sources at 90\,GHz and shorter monitoring period which also 
shortens the timescales obtained at 90\,GHz. We studied the differences 
between the source classes at 22 and 37\,GHz and at both 
frequencies the classes do not differ significantly from each other if the 
observational uncorrected timescales are considered. When we study the 
redshift corrected timescales we see significant differences between 
the BLOs and LPQs at 37\,GHz and in addition there is strong 
indication that also the HPQs and BLOs differ. Similar results were obtained 
in Paper I, when periodogram timescales were studied. 

We were also interested in sources in which the timescale is present 
during most of the flux curve. By visually extracting the duration of the 
timescale from the wavelet plot, we calculated how much of the total 
monitoring time it has been present. In 27 cases the timescale was present 
at least 90\% of the total observing time. Seven of these, 
however, were 
cases in which the timescale had not repeated twice during the period 
it was visible.

\subsection{Flare timescales}
In addition to long-term trends we determined a flare timescale which 
describes the duration or rise and decay times of flares.
These timescales are seen in the wavelet transform at the times 
of the outbursts. We determined such a timescale in 53 cases. 
They are also tabulated in Table \ref{table:timescales}.  
We have taken into account cases for which a similar timescale seems 
to be present during more than one flare 
and therefore represented a typical flare timescale for the source. The 
values we obtained vary between 0.34--2.17 years at 22 and 37\,GHz. 
The average value at 22\,GHz is 1.16 years and at 37\,GHz 
it is 1.03 years. At 90\,GHz, there were only 
6 sources which showed repeating flares and the timescales were shorter, 
varying between 0.34 and 0.87 years, with an average of 0.69 years. 

\subsection{Individual sources}
Even though none of the sources studied here seems to show a strict periodicity 
in radio frequencies, many of them show quasi-periodic behaviour. We are 
interested in sources that have a variability timescale lasting four cycles 
in at least two of the frequency bands. 
There are eight such sources and their properties are discussed in more detail 
below. The wavelet plots for these sources at 37\,GHz are shown in Figs. \ref{0133}--\ref{1749} (only available in electronic form via http://www.aanda.org).

\onlfig{2}{
\begin{figure*}
\includegraphics[height=17cm, width=11.5cm, angle=-90]{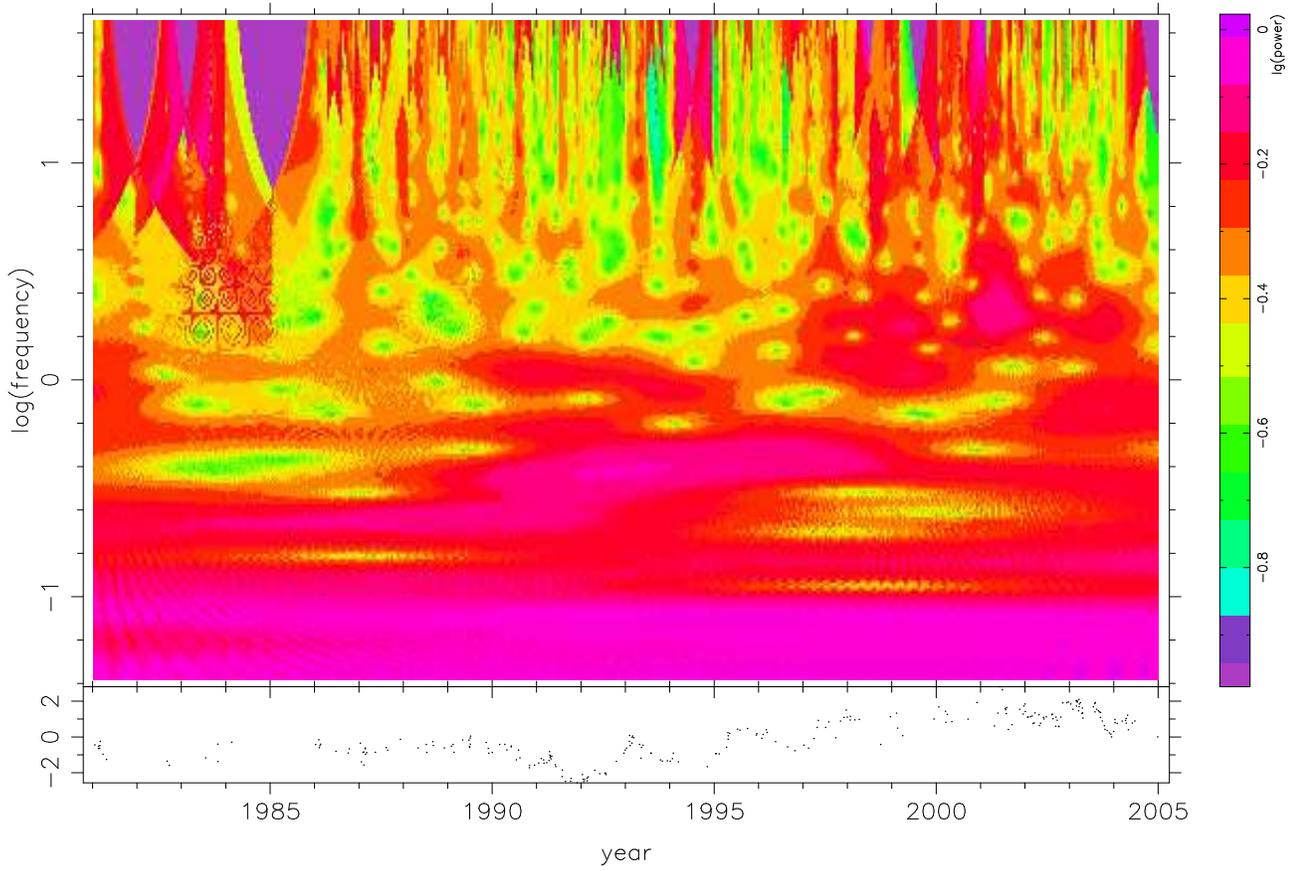}
\caption{Wavelet transform of the source \object{DA 55} (\object{0133+476}) at 37\,GHz.}
\label{0133}
\end{figure*}
}

\onlfig{3}{
\begin{figure*}
\includegraphics[height=17cm, width=11.5cm, angle=-90]{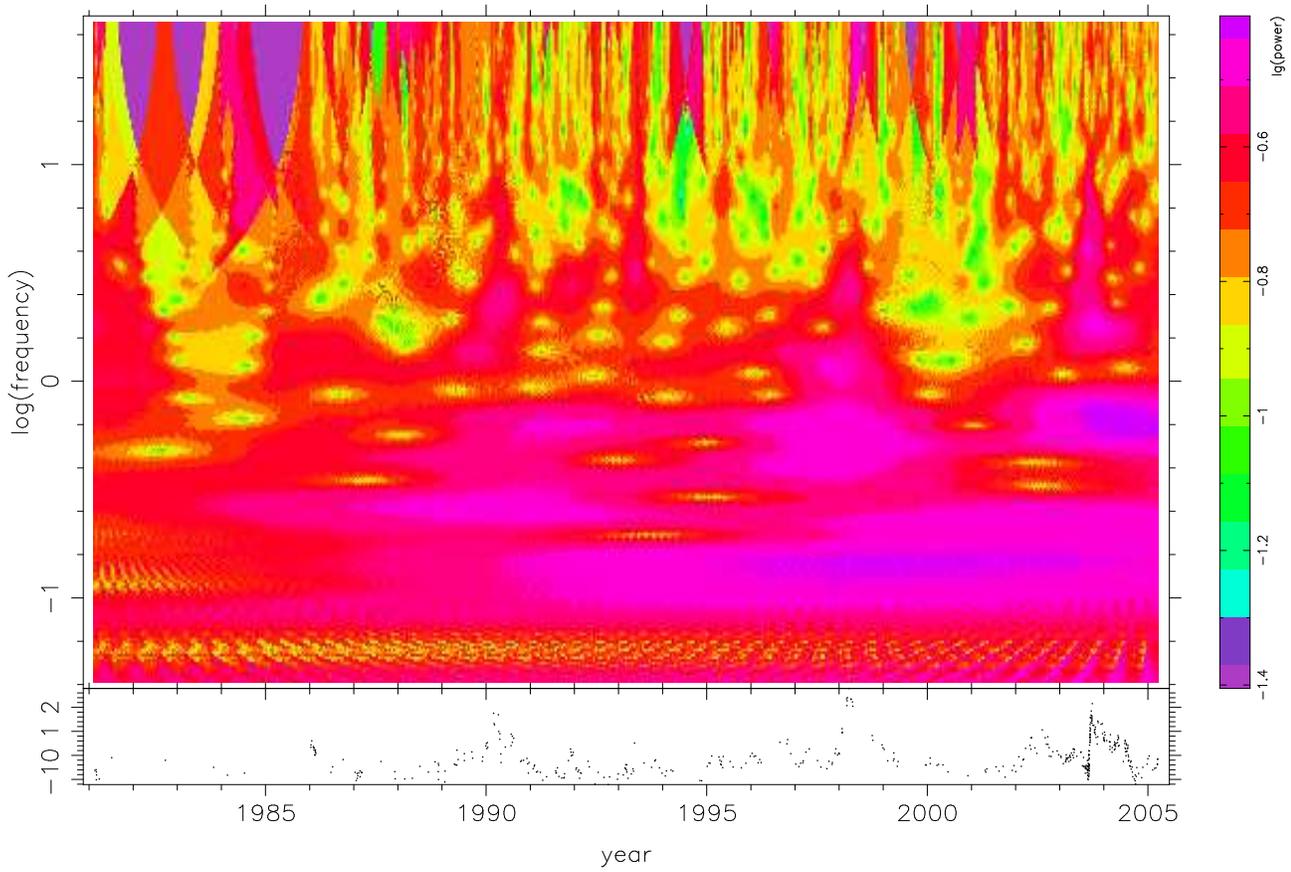}
\caption{Wavelet transform of the source \object{3C 120} (\object{0430+052}) at 37\,GHz.}
\label{0430}
\end{figure*}
}

\onlfig{4}{
\begin{figure*}
\includegraphics[height=17cm, width=11.5cm, angle=-90]{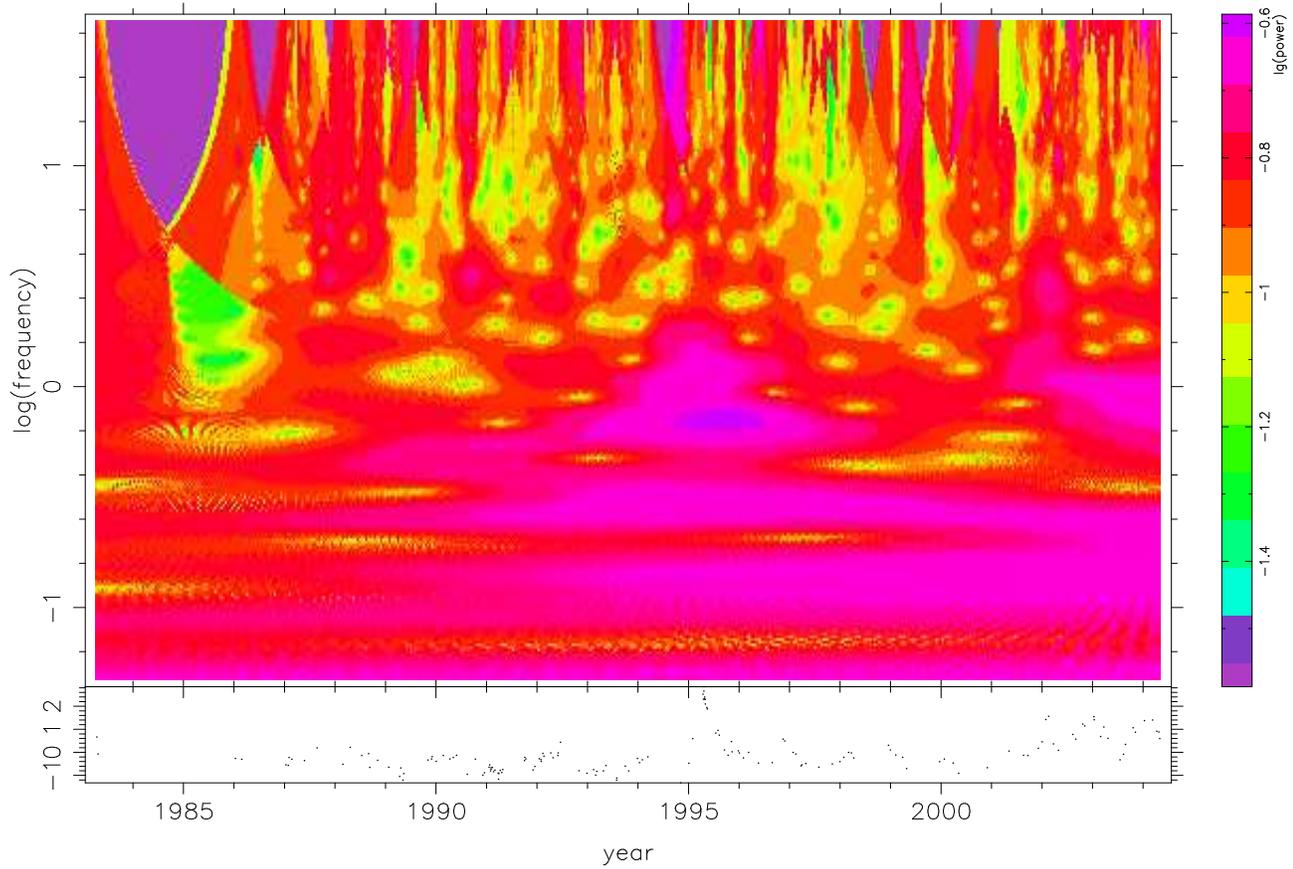}
\caption{Wavelet transform of the source \object{0736+017} at 37\,GHz.}
\label{0736}
\end{figure*}
}

\onlfig{5}{
\begin{figure*}
\includegraphics[height=17cm, width=11.5cm, angle=-90]{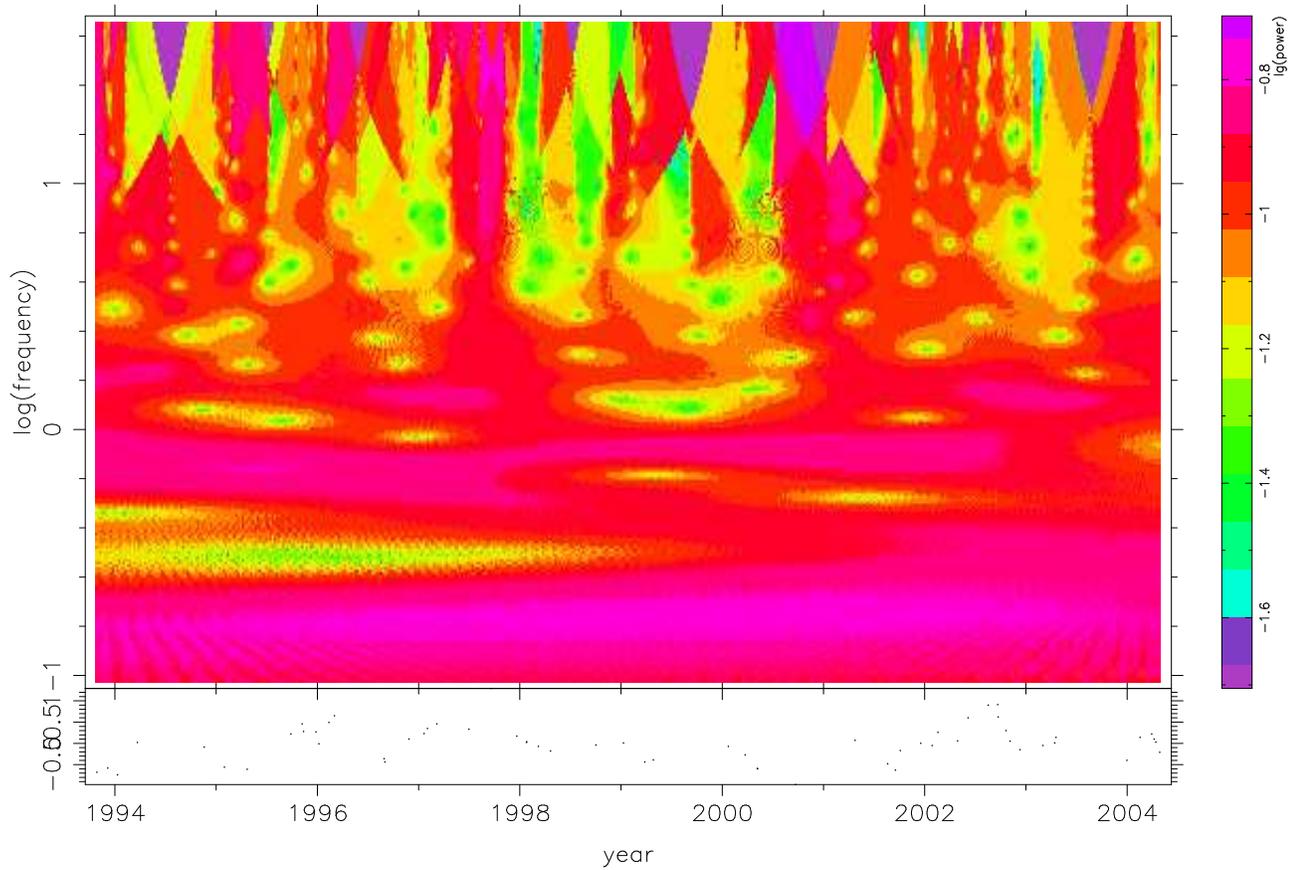}
\caption{Wavelet transform of the source \object{OJ 248} (\object{0827+243}) at 37\,GHz.}
\label{0827}
\end{figure*}
}

\onlfig{6}{
\begin{figure*}
\includegraphics[height=17cm, width=11.5cm, angle=-90]{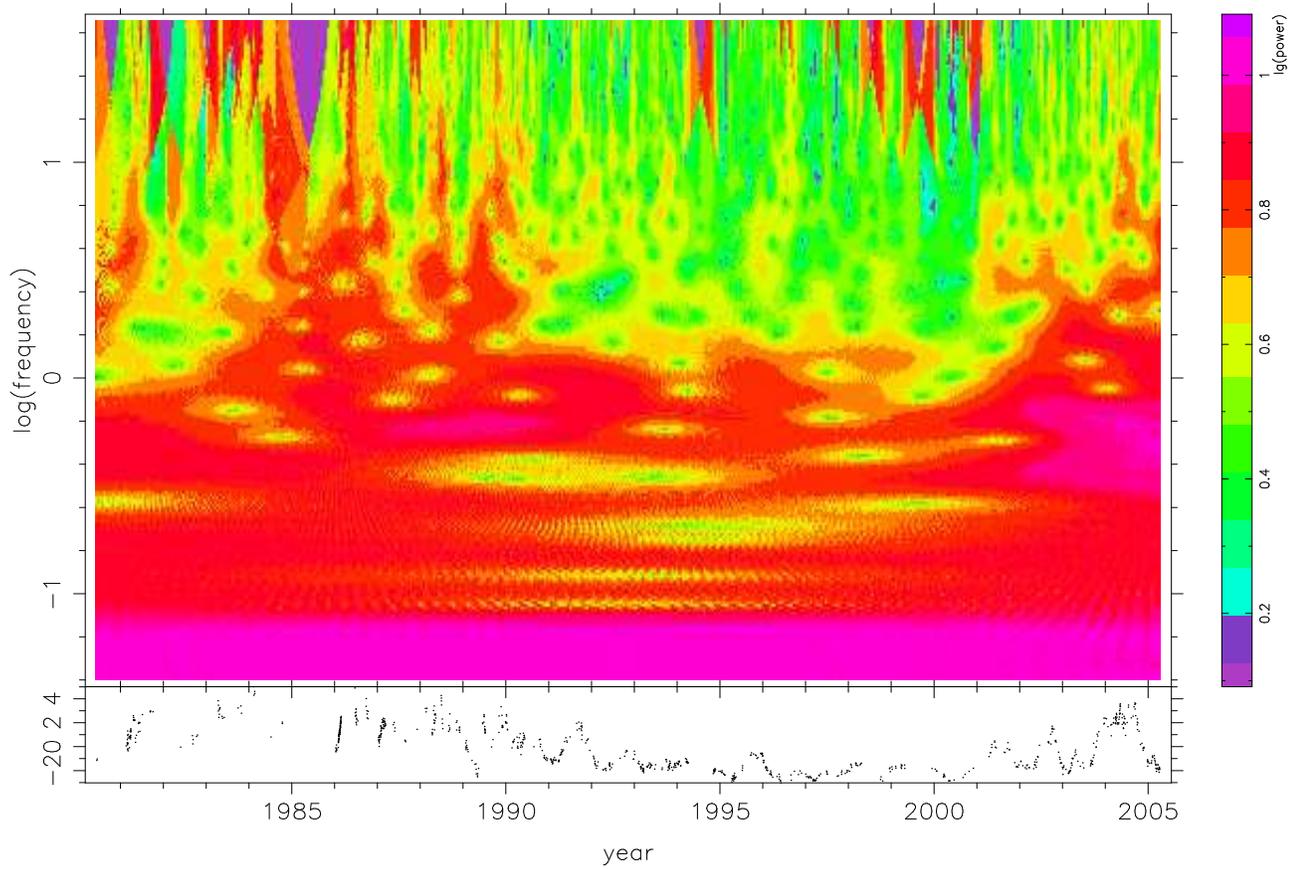}
\caption{Wavelet transform of the source \object{OJ 287} (\object{0851+202}) at 37\,GHz.}
\label{0851}
\end{figure*}
}

\onlfig{7}{
\begin{figure*}
\includegraphics[height=17cm, width=11.5cm, angle=-90]{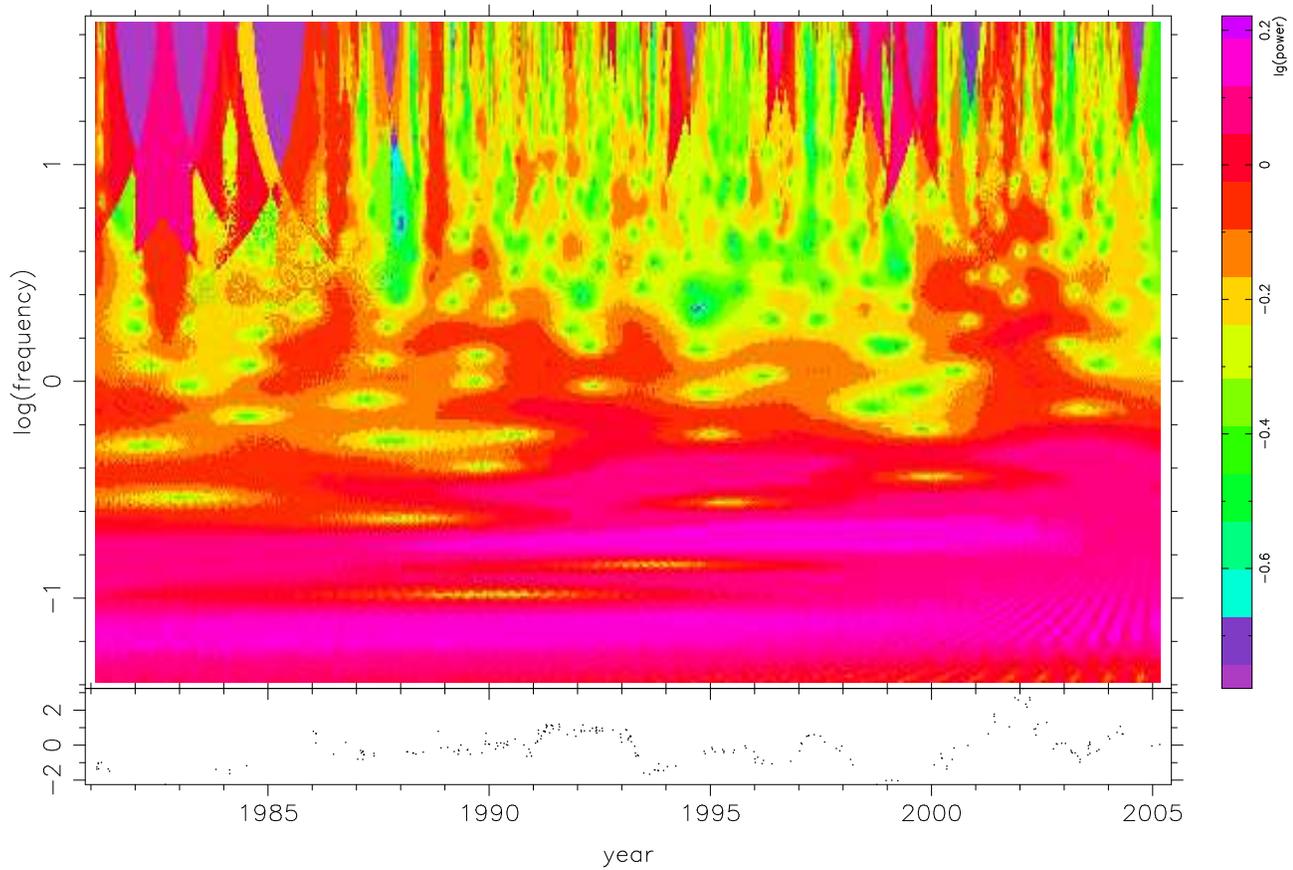}
\caption{Wavelet transform of the source \object{OL 093} (\object{1055+018}) at 37\,GHz.}
\label{1055}
\end{figure*}
}

\onlfig{8}{
\begin{figure*}
\includegraphics[height=17cm, width=11.5cm, angle=-90]{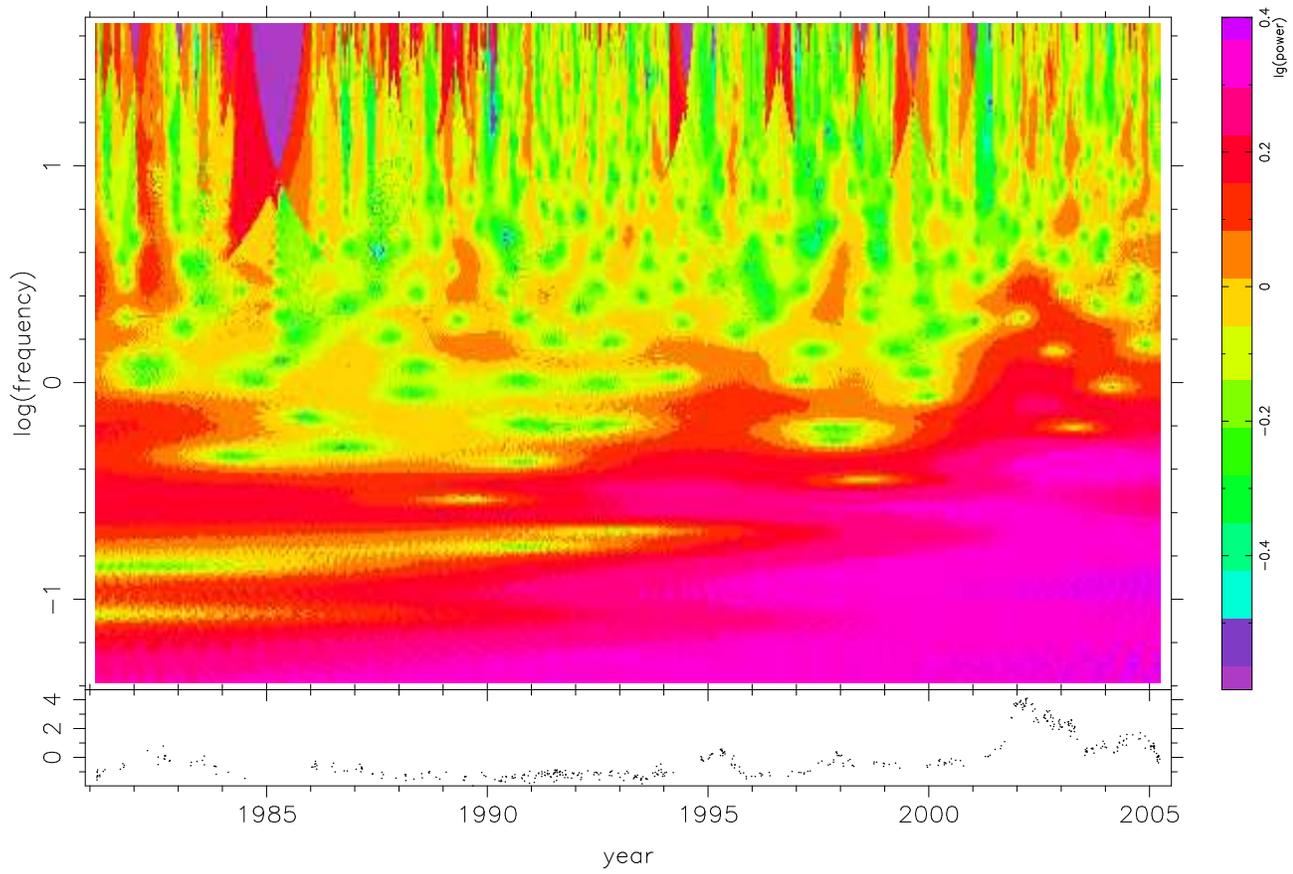}
\caption{Wavelet transform of the source \object{4C 38.41} (\object{1633+382}) at 37\,GHz.}
\label{1633}
\end{figure*}
}

\onlfig{9}{
\begin{figure*}
\includegraphics[height=17cm, width=11.5cm, angle=-90]{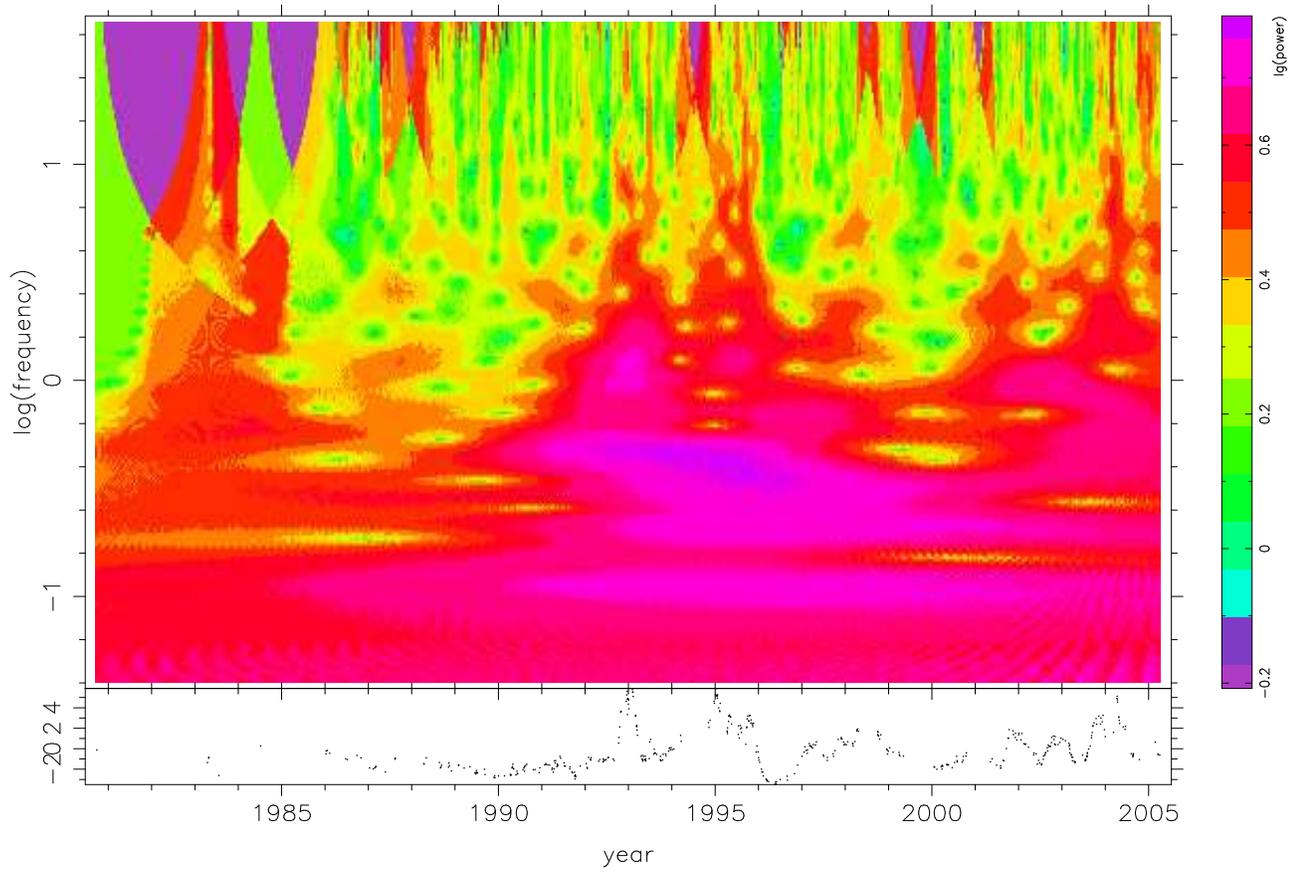}
\caption{Wavelet transform of the source \object{PKS 1749+096} at 37\,GHz.}
\label{1749}
\end{figure*}
}

\textbf{\textit{\object{DA 55} (\object{0133+476})}}:
This HPQ source has been monitored in Mets\"ahovi for over 20 years at 
22 and 37\,GHz and for almost 15 years at 90\,GHz. It has shown 
high apparent pattern velocity of $18c$ in the jet \citep{piner07} and 
has been also a target of many other studies using Very Long 
Baseline Interferometry (VLBI) observations 
\citep[e.g.][]{lister01, kellermann04, lister05}. It is also included in 
the wavelet study of \cite{kelly03} which uses the UMRAO 4.8, 8, 14.5\,GHz 
monitoring data. They found a timescale of 2.3 years for the source at 
14.5\,GHz. This is very close to the timescale of 2.2 years which we obtained 
for the source at both 22 and 37\,GHz. At 90\,GHz we only found a flare 
timescale of 0.7 years for the source. The long-term timescale in our 
analysis has a rising trend so that it is shortening towards a peak of 
a large flare. The timescale is present only for about 40\% of the flux curve 
at both 22 and 37\,GHz but still has repeated 4.4 times at 
both of the frequency bands. In Paper I we could not find any DCF or 
periodogram timescale for the source. When the flux curve is examined we 
can clearly see that the timescale obtained here is related to a period 
when the source had many similar flares and the total flux level rises. This 
also explains the rising trend in the wavelet plot which is caused by 
more frequent individual flares towards the peak of the larger event.

\textbf{\textit{\object{3C 120} (\object{0430+052})}}: 
This radio galaxy has been monitored in Mets\"ahovi for over 20 years at 
22 and 37\,GHz and for 15 years at 90\,GHz. It is a well-studied 
object at all wavelengths and has been a target of many multiwavelength 
campaigns. Its radio jet has been studied and modelled using VLBI 
observations by several authors 
\citep[e.g.][]{walker01, gomez01, hardee05}. Our wavelet analyses reveal 
a timescale of 4.3 years at both 22 and 37\,GHz. At 22\,GHz the timescale 
has been present for 85\% of the total observing period lasting for  
4.4 cycles. At 37\,GHz it has been present for 84\% of the time and 
persisted for 4.6 cycles. In Paper I the DCF analyses showed a timescale 
of 4.2 years at 22\,GHz which is very close to what we obtained here. 
Both frequencies also show flare timescales of 1.4 years and in addition 
0.5 years 
at 22\,GHz and 0.3 years at 37\,GHz. These timescales are also seen in the 
DCF and SF analyses of Paper I. At 90\,GHz the timescale obtained is 
2.7 years and it has been present for 73\% of the time and continued for 4 cycles. This timescale is also very close to the DCF timescale of 
2.9 years.

Visual inspection of the flux curve at 37\,GHz shows that indeed many of 
the larger flares have approximately 4.5 to 5 years between them but 
there are also additional flares in between.
The last big flare included in our analysis at 37\,GHz peaked in the autumn 
of 2003. The next big flare was observed at the end of 2006 with an 
interval of 3 years between the flares. This indicates that the source is not 
strictly periodic but has shown a characteristic timescale of 4.3 years.

\textbf{\textit{\object{0736+017}}}:
This HPQ source was not detected by EGRET but has been listed as a 
possible $\gamma-$ray source to be detected with the new satellite 
missions AGILE and GLAST \citep{bach07}. In Mets\"ahovi it has been 
monitored for over 20 years at 22 and 37\,GHz and for 11 years at 90\,GHz.
It exhibits rapid variability and in the optical domain it has been studied 
by several authors \citep[e.g.][]{clements03, ramirez04}. The wavelet 
plot has a complex structure at all the frequency bands and we found 
multiple timescales for the source. The most significant and long-lasting 
ones are 2.7, 3.4 and 1.7 years at 22, 37 and 90\,GHz, respectively. 
The timescales have repeated themselves 4 times at all the frequency bands 
and are also present for about half of the flux curve. The DCF analysis of 
Paper I shows very similar timescales of 2.8 years at 22 and 37\,GHz and 1.2 
years at 90\,GHz. At 22 and 37\,GHz there are multiple timescales present 
also in the DCF which confirms the complex variability of this source. 
Therefore it is also difficult to predict when the next flares could be 
expected. 
 
\textbf{\textit{\object{OJ 248} (\object{0827+243})}}:
This radio quasar was found by EGRET to be a bright $\gamma-$ray source. It has been monitored at 
Mets\"ahovi for about 11 years both at 22 and 37\,GHz. It has also 
been a target of multiwavelength campaigns \citep[e.g.][]{bach07} and 
has shown extremely fast apparent pattern velocity exceeding 25c 
in the jet \citep{piner06}. The kiloparsec-scale jet is studied in 
more detail in \cite{jorstad04}. We found a timescale of 1.4 years 
at both frequency bands but it appears quite weak in strength and is 
classified as a weak timescale at 22\,GHz.
At 22\,GHz 
it has been present for 54\% of the time and is discontinuous. It 
repeated 4.2 times during the time it has been present. At 37\,GHz 
the timescale has been present for 77\% of the time and has a shortening 
trend changing the timescale from 1.4 to 1.1 years. The timescale
repeated 6.3 times. In Paper I we measured
timescales of 1.7 
and 1.4 years at 22 and 37\,GHz respectively from the DCF analyses. 

\textbf{\textit{\object{OJ 287} (\object{0851+202})}}:
This BLO type object is one of the most studied AGNs. In Mets\"ahovi it has 
been monitored for over 25 years at 22 and 37\,GHz 
and at 90\,GHz for 15 years. 
In the optical there is strong evidence of a period of about 12 years 
which is explained by several binary black hole models 
\citep[e.g.][]{sillanpaa88, lehto96, valtaoja00, valtonen08}. 
We can neither confirm nor disprove the existence of the 
same period at 
our frequency bands because the monitoring time is too short compared to the 
optical historical light curve which is over 100 years long.
The source has also been studied with wavelets by \cite{hughes98}. 
They used the continuous wavelet transform to study more than 20 years 
of data obtained in the University of Michigan Radio Observatory (UMRAO) at 
4.8, 8 and 14.5\,GHz. They found a persistent timescale of 1.66 years in 
the total flux and polarisation and another timescale of 1.12 years 
dominating the activity in the 1980s. They explained the modulations with a 
shock-in-jet model in which the permanent timescale is associated with 
the quiescent jet and the shorter timescale is due to the passage of a 
shock.

We found a timescale of 1.4 years at 22 and 90\,GHz and 1.7 years at 
37\,GHz. Both of these can be considered to be very similar to the ones 
obtained by \cite{hughes98}, given the ambiguity in determining the exact time 
scale in our analysis. A very interesting phenomenon seen in the wavelet 
transform is that the timescale either weakened or totally disappeared 
between 1993 and 2000. Due to this the timescale is 
present only for 2/3 of the time in all the frequency bands. It can be 
clearly seen in the flux density curve in Fig. \ref{0851} that the flux density level is 
generally lower during that time, even though smaller flares occur 
continuosly. It is also possible to see in the wavelet plot that the number of rapid flares is lower during that period than in the more active states.
Nevertheless the timescale repeated 10.2 times at 22\,GHz,
9.6 times at 37\,GHz and 7.2 times at 90\,GHz. All the timescales have a 
corresponding one in the DCF analysis of Paper I within 6 months of the one 
obtained from wavelet analysis. When studying the flare timescales we noticed 
timescales of 0.4 and 0.3 years at 22 and 37\,GHz, respectively. 
These agree with timescales from the SF 
analysis in Paper I. It is difficult to associate the timescales with 
individual flares because the source exhibits continuous variability and is 
almost never in a quiescent state.

\textbf{\textit{\object{OL 093} (\object{1055+018})}}:
This HPQ source has been observed in Mets\"ahovi for over 20 years 
at 22 and 37\,GHz and for almost 13 years at 90\,GHz. The jet of this 
source is suggested to have a spine-sheath structure 
\citep{attridge99,pushkarev05}. We obtained a timescale of 4.3 years 
for the source at both of the 22 and 37\,GHz frequencies. The timescale 
repeated 4.4 times and is present for 85 and 80\% of 
the flux curve at 22 and 37\,GHz, respectively. Very similar timescales 
are measured in Paper I, where a DCF timescale of 4.7 and 4.4 years are 
obtained at 22 and 37\,GHz, respectively. At 90\,GHz we obtain a timescale 
of 2.7 years but it repeated only 2.6 times. When examining the 
flux curve at 37\,GHz we see that indeed there are many flares with 
approximately 4--5 years between them. The last flare included in the 
wavelet analysis peaked in 2002 and another large flare was observed in 
2007, producing a time interval of 5 years between the flares. In addition 
there was a smaller flare peaking in 2004 which shows that even though 
the timescale of 4.3 years describes the larger events quite well, 
there is still more complex variability present in this source.

\textbf{\textit{\object{4C 38.41} (\object{1633+382})}}:
This is another EGRET-detected $\gamma-$ray bright quasar which 
has been a target of many different observing programs 
\citep[e.g.][]{katajainen00, jorstad01, lister05, bach07}. In 
Mets\"ahovi it has been monitored for over 20 years at 22 and 37\,GHz. 
It is also included in the wavelet study of \cite{kelly03} 
which used the UMRAO data at 4.8, 8 and 14.5\,GHz. In their analysis 
no timescale was found for the source. We found a timescale of 
3.4 years in both frequency bands but the wavelet plot has a complex 
structure and the variability timescale is weaker in the late 1980s 
and early 1990s. The timescale has been present for almost 3/4 of the 
flux curve in both frequency bands and 
repeated 4.8 and 5.1 times at the 22 and 37\,GHz 
frequencies, respectively. The timescale is not 
seen in DCF analysis of Paper I, probably because even though there are some 
flares with 3.5 years between them, it is not a timescale clearly seen to 
repeat in the flux curve.

\textbf{\textit{\object{PKS 1749+096}}}:
This BLO type object has been monitored in Mets\"ahovi for 20 years at 
22\,GHz and 25 years at 37\,GHz. At 90\,GHz the data sets are 
14 years long. The source has been a target of many Very Long Baseline 
observations \citep[e.g.][]{homan01, wiik01, lister05, piner07}.
We found a timescale of 2.7 years in all the frequency bands but the 
structure of the wavelet plot is very complex and the timescale 
is increasing in time from 1.7 to 2.7 years. At 90\,GHz it also 
repeated only 3.4 times while at 22\,GHz 4.4 and at 37\,GHz 
5 times. At all frequency bands the timescale has also been present for less 
than 70\% of the time. Nevertheless 
timescales of 2.8 years at 22 and 37\,GHz and 2.3 years at 90\,GHz 
are also seen in the DCF analysis of 
Paper I. In addition we found flare timescales of 0.4 and 0.5 years at 
22 and 37\,GHz, respectively. Both are also seen within 0.2 years in the 
SF analysis of Paper I. 
The source exhibits rapid variability and has many 
large flares in the flux curve. Indeed many of the flares seem to be within 
2.7 years of each other but there is also other activity present making 
the analysis complex.

\section{Discussion} \label{sect:discussion}
Our interest in using wavelets to study the timescales arose from the 
results of Paper I, which showed that many of the sources have 
changed their behaviour during the monitoring time and the timescales 
have changed over the years. A useful property of wavelets is that 
they show also when the timescale has been present and how it has 
changed. Many studies on individual sources have claimed the sources 
to be periodic or quasi-periodic based on results from Fourier-based 
methods. Our results show that only a very small number of sources 
actually show persistent timescales lasting over the total monitoring 
time (18 sources in which the timescale had lasted for over 90\% of the total monitoring time). 
A good example is the source \object{1156+295} shown in Fig. 
\ref{example}. The wavelet timescale of 3.4 years is seen to be present 
only in the latter half of the flux curve. In Fig. 11 of Paper I the DCF and 
Lomb-Scargle periodogram analyses are shown for the same source. The DCF 
gives a timescale of 3.5 years and the Lomb-Scargle periodogram a timescale 
of 3.3 years which are very close to the timescale obtained with wavelets. 
With these methods, however,  it was impossible to 
see that the source had changed its behaviour in the mid 1990s and the 
timescale was only present for the latter half of the monitoring period.
In eight cases the timescale either disappeared or weakened 
at some point during the monitoring period and re-appeared later on.
These changes cannot be detected by Fourier-based methods. Therefore we 
feel that wavelets should be used more when characteristic timescales and 
quasi-periodicities are studied.

In many cases the timescale also changed slowly over time. In 17 cases we 
saw a rising trend in the wavelet plot, which means a shortening timescale.
These were often connected to long-lasting flares which showed shorter 
timescales towards the peak of the flare. This could indicate that 
there are more disturbances in the jet when a flare is growing and 
more shocks are developing in the jet, making the timescale shorter. 
Also, in six cases we detected a declining trend in the wavelet plot, indicating 
an increasing timescale. In these cases the activity is becoming less frequent 
and flares do not occur as often as earlier.

\begin{figure}
\resizebox{\hsize}{!}{\includegraphics{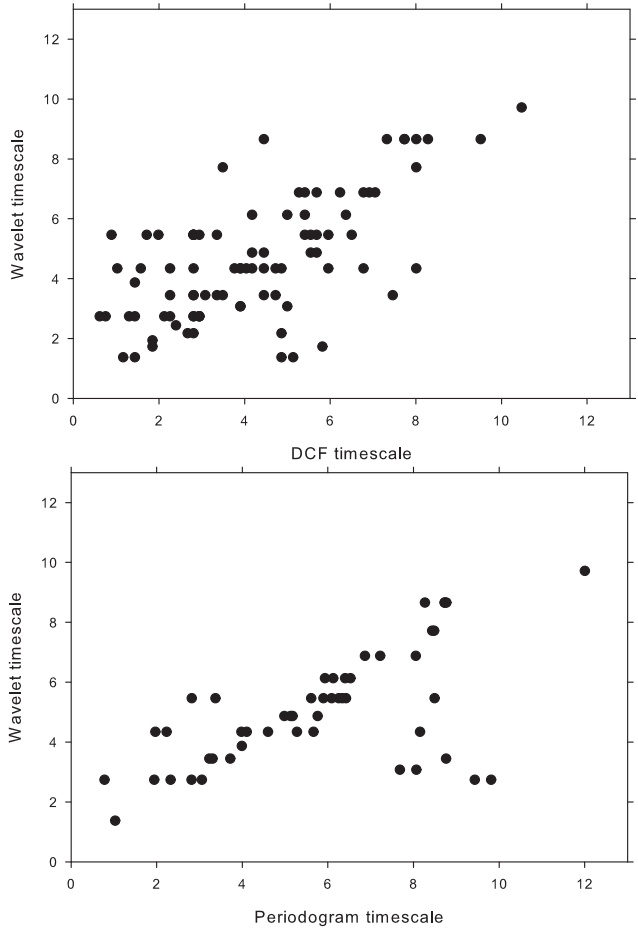}}
\caption{Upper panel: Long-term wavelet timescale against the DCF timescale from Paper I. Lower panel: The same wavelet timescale against the Lomb-Scargle periodogram timescale from Paper I.}
\label{wavelet_dcf_ls}
\end{figure}

In Paper I we also found that the traditional Fourier-based methods have many 
problems when non-sinusoid flux curves 
are studied. In our sample this is 
the case for all sources. Especially with Lomb-Scargle periodogram, 
we obtain 
spurious timescales and other methods are needed to confirm the timescales. 
The DCF seemed slightly more reliable but it also has no means of showing 
if the timescale is discontinuous or changes in time. We 
compared our results from the wavelet analysis 
with those obtained in Paper I. The wavelet timescales 
are plotted against the DCF and periodogram timescales in Fig. 
\ref{wavelet_dcf_ls}. We can see a very good correspondence between the 
different timescales. We also calculated the Spearman rank correlation 
and between wavelet and DCF timescales we obtain a correlation of $r=0.58$ ($p<0.0001$). Similarly 
between wavelet and periodogram timescales we obtain $r=0.55$ ($p<0.0001$).
The correlation between the wavelet and periodogram timescales is 
affected by sources that have longer periodogram timescales of over 7 
years but shorter wavelet timescales of 2 to 5 years. In almost all cases 
the periodogram analysis also revealed another
timescale, similar to the wavelet timescale, but in our study 
we only used the most significant timescale
from the periodogram analysis. Only for one source 
(\object{0736+017}) did the periodogram analysis give only long timescales of 
over 8 years whereas 
the wavelet analysis gave short 2.7 and 3.4 years 
timescales. This source, 
however, has a complex structure in the wavelet plot (as seen in Fig. \ref{0736}) and multiple timescales are seen.

The average timescales of wavelet and DCF analyses are also very similar 
with the difference being less than 6 months at all frequency bands.
The averages of the periodogram analyses are slightly higher. The results show 
that both wavelet and DCF analyses give similar results and can be used 
to determine characteristic timescales. The advantage of wavelets is that 
we can also take into account how long a certain timescale has been present 
in the flux curve.

In our sample we had nine sources for which \cite{kelly03} found variability timescales 
at some of the UMRAO frequencies. For one of those (\object{1823+568}) we were not able to find 
any variability timescale in our analyses. In almost all cases we were able to find a 
similar timescale within 0.5 years of the timescale reported at one of the low 
frequencies. Only in one case was the timescale totally different (\object{1641+399}) and in 
one case (\object{0212+735} at 37\,GHz) there was a difference of 0.7 years between the timescales. In three
cases it was the timescale we refer to as the flare timescale that was 
similar to the lower frequency timescale. The similarity between the results from 
the two analyses supports use of the method.

Even though we found no strict periodicities in the radio behaviour
of any of our sources, either in Paper I or in this paper, we wanted to
investigate the possibility of ``predicting'' upcoming active states in
these sources. Many of these sources have changed 
their behaviour during the 20 years of observations (for example 
\object{1156+295} in Fig. \ref{example}), thus such predictions would be 
based purely on their statistical behaviour. This attempt was partly motivated by
the multifrequency support observations we participate in:
it would be immensely useful to be able to prioritise 
those sources that are likely to enter an active state fairly soon. 
In particular, we were interested to see if we can predict which sources 
would be likely to flare during the early operation of the 
upcoming gamma-ray satellite GLAST. Our
earlier work \citep[e.g.][]{tornikoski02, lahteenmaki03} has shown
that strong gamma-ray activity usually corresponds to a growing flare 
in the radio regime. Therefore sources that are active in the radio 
domain are potential sources to be detected by GLAST.

Thus, we wanted to see which sources in our sample would be good candidates
for strong flaring within the timeframe 2008--2009, excluding a set of sources
that have almost continuous and complex
variability at 37\,GHz, such as  \object{OJ 287} and \object{BL Lac}.
From the wavelet timescales at
37\,GHz, combined with the timescale analysis from Paper I and with visual
examination of light curves
until the end of 2007, we came up with a list of six sources that
according to the wavelet analysis and at least one other method
used in Paper I show semi-periodicity with such timescales that
a strong radio flare could be anticipated to occur within 2008--2009.
The sources include one GAL
type object (\object{0007+106}), two HPQs (\object{0234+285},
\object{1156+295}), two LPQs (\object{0333+321}, \object{2145+067}) and one
BLO (\object{1749+096}).

\section{Conclusions} \label{sect:conclusions}
We studied a sample of 80 sources with the continuous wavelet transform
using data at 22, 37 and 90\,GHz. Our aim was to study the variability 
behaviour of the sources and also to better 
understand the method and to compare it 
with Fourier-based methods. We found no clear periodicities in the 
sources. Instead in most of the sources the 
timescales appeared only for a short time in the total monitoring period or 
changed over the years. These kind of properties are not revealed with 
Fourier-based methods and therefore wavelets should be used more when 
quasi-periodicities are studied. In our sample only 1/4 had a timescale 
that had been present for at least 4 cycles. In eight of these sources the timescale was present in more than one frequency band. This shows that the sources are never strictly periodic and conclusions about periodicities in the radio frequencies should be drawn with caution.

The average timescale from the wavelet analysis is 4.5 years which is the 
same as was obtained in the DCF analysis of Paper I. This timescale should 
not be affected by the length of the monitoring period because we used 
all the sources in which the timescale had repeated at least twice in the 
analysis and therefore longer timescales also are included in the 
calculations. The weakness of 
wavelets is that we cannot detect the timescale with the same accuracy as 
with DCF or periodograms and therefore both type of methods should be 
used when periodicities or quasi-periodicities are searched for. The wavelets 
should be used to verify whether the timescale is persistent or short-lived. 
If a persistent timescale is found, the Fourier methods can be used to 
accurately determine the timescale. In our analysis where the characteristic 
timescales are studied, the accuracy of wavelets is enough to determine 
the approximate timescale.

\begin{acknowledgements}
We acknowledge the support of the Academy of Finland (project numbers 212656
and 205793).
\end{acknowledgements}

\bibliographystyle{/home/tho/texmf/tex/aa-package/bibtex/aa}
\bibliography{/home/tho/texmf/tex/aa-package/bibtex/thbib_wave}

\longtabL{1}{
\begin{landscape}
\small{
\begin{longtable}{lll|cccc|cccc|cccc}
\caption{\label{table:timescales}Timescales from the wavelet analyses. For each source the long term timescale is given in years and the number of cycles it has repeated. In addition flare timescale and total monitoring time are shown. If more than one timescale is given the most significant one is placed first.}\\
\hline
\hline
 	  & 	  	  & 	  	  & 	 \multicolumn{4}{c}{22\,GHz} 	   	  	   	  	   	  	  &  \multicolumn{4}{c}{37\,GHz}	  & 	 \multicolumn{4}{c}{90\,GHz}	  \\
B1950 	  & 	 Other 	  & 	 Class 	  & 	 monitoring 	  & 	 time 	  & 	 cycles 	  & 	 flare 	  & 	 monitoring 	  & 	 time	  & 	 cycles 	  & 	 flare 	  & 	 monitoring 	  & 	 time 	  & 	 cycles 	  & 	 flare 	  \\
 name 	  & 	 name 	  & 	  	  & 	 time [yr]	  & 	 scale [yr]	  & 	  	  & 	 scale [yr]	  & 	 time [yr]	  & 	 scale [yr]  & 	  	  & 	 scale [yr] 	  & 	 time [yr]	  & 	 scale [yr]	  & 	  	  & 	 scale [yr] 	  \\
\hline 
\endfirsthead
\caption{Continued.}\\
\hline
\hline
 	  & 	  	  & 	  	  & 	 \multicolumn{4}{c}{22\,GHz} 	   	  	   	  	   	  	  &  \multicolumn{4}{c}{37\,GHz}	  & 	 \multicolumn{4}{c}{90\,GHz}	  \\
B1950 	  & 	 Other 	  & 	 Class 	  & 	 monitoring 	  & 	 time	  & 	 cycles 	  & 	 flare 	  & 	 monitoring 	  & 	 time	  & 	 cycles 	  & 	 flare 	  & 	 monitoring 	  & 	 time	  & 	 cycles 	  & 	 flare 	  \\
 name 	  & 	 name 	  & 	  	  & 	 time [yr]	  & 	 scale [yr]	  & 	  	  & 	 scale [yr]	  & 	 time [yr]	  & 	 scale [yr]	  & 	  	  & 	 scale [yr]	  & 	 time [yr]	  & 	 scale [yr]	  & 	  	  & 	 scale [yr]	  \\
\hline 
\endhead
\hline
\endfoot
 \object{0007$+$106} 	  & 	 III ZW 2 	  & 	 GAL 	  & 	 19.41 	  & 	 4.9 	  & 	 3.5 	  & 	 0.8 	  & 	 19.23 	  & 	 $4.9^r$ 	  & 	 3.0 	  & 	 0.9 	  & 	 N 	  & 	 N 	  & 	 N 	  & 	 N 	  \\
 \object{0016$+$731} 	  & 	  	  & 	 LPQ 	  & 	 11.87 	  & 	 - 	  & 	 - 	  & 	 0.3 	  & 	 10.81 	  & 	 - 	  & 	 - 	  & 	 - 	  & 	 N 	  & 	 N 	  & 	 N 	  & 	 N 	  \\
 \object{0106$+$013} 	  & 	 OC 012 	  & 	 HPQ 	  & 	 22.27 	  & 	 6.9 	  & 	 1.4 	  & 	 1.7 	  & 	 24.15 	  & 	 8.7 	  & 	 1.6 	  & 	 2.2 	  & 	 N 	  & 	 N 	  & 	 N 	  & 	 N 	  \\
 \object{0109$+$224} 	  & 	 S2 0109+22 	  & 	 BLO 	  & 	 19.33 	  & 	 8.7 	  & 	 1.6 	  & 	 - 	  & 	 20.66 	  & 	 $2.7^r$ 	  & 	 5.0 	  & 	 1.4 	  & 	 N 	  & 	 N 	  & 	 N 	  & 	 N 	  \\
 \object{0133$+$476} 	  & 	 DA 55 	  & 	 HPQ 	  & 	 22.43 	  & 	 $2.2^r$ 	  & 	 4.4 	  & 	 - 	  & 	 24.17 	  & 	 $2.2^r$ 	  & 	 4.4 	  & 	 - 	  & 	 14.92 	  & 	 - 	  & 	 - 	  & 	 0.7 	  \\
 \object{0149$+$218} 	  & 	  	  & 	 LPQ 	  & 	 15.89 	  & 	 - 	  & 	 - 	  & 	 - 	  & 	 16.69 	  & 	 - 	  & 	 - 	  & 	 - 	  & 	 N 	  & 	 N 	  & 	 N 	  & 	 N 	  \\
 \object{0202$+$149} 	  & 	 4C 15.05 	  & 	 HPQ 	  & 	 19.93 	  & 	 $4.3^c$ 	  & 	 2.5 	  & 	 0.9 	  & 	 20.66 	  & 	 $4.3^c$ 	  & 	 4.4 	  & 	 1.1 	  & 	 N 	  & 	 N 	  & 	 N 	  & 	 N 	  \\
 \object{0212$+$735} 	  & 	  	  & 	 HPQ 	  & 	 15.97 	  & 	 - 	  & 	 - 	  & 	 - 	  & 	 16.74 	  & 	 $3.4^f$ 	  & 	 3.7 	  & 	 - 	  & 	 N 	  & 	 N 	  & 	 N 	  & 	 N 	  \\
 \object{0224$+$671} 	  & 	  	  & 	 QSO 	  & 	 13.37 	  & 	 2.7 	  & 	 4.6 	  & 	 - 	  & 	 14.87 	  & 	 $2.2^c$ 	  & 	 3.8 	  & 	 - 	  & 	 N 	  & 	 N 	  & 	 N 	  & 	 N 	  \\
 \object{0234$+$285} 	  & 	 4C 28.07 	  & 	 HPQ 	  & 	 16.17 	  & 	 - 	  & 	 - 	  & 	 - 	  & 	 15.08 	  & 	 2.4 	  & 	 4.5 	  & 	 - 	  & 	 10.26 	  & 	 4.3 	  & 	 1.6 	  & 	 - 	  \\
 \object{0235$+$164} 	  & 	  	  & 	 BLO 	  & 	 22.44 	  & 	 $5.5^c$ 	  & 	 3.5 	  & 	 $1.7,0.3^c$ 	  & 	 24.11 	  & 	 - 	  & 	 - 	  & 	 0.3 	  & 	 12.31 	  & 	 1.9 	  & 	 2.3 	  & 	 - 	  \\
 \object{0248$+$430} 	  & 	  	  & 	 LPQ 	  & 	 17.59 	  & 	 - 	  & 	 - 	  & 	 - 	  & 	 20.73 	  & 	 - 	  & 	 - 	  & 	 - 	  & 	 N 	  & 	 N 	  & 	 N 	  & 	 N 	  \\
 \object{0316$+$413} 	  & 	 3C 84 	  & 	 GAL 	  & 	 22.45 	  & 	 - 	  & 	 - 	  & 	 - 	  & 	 25.46 	  & 	 - 	  & 	 - 	  & 	 - 	  & 	 14.95 	  & 	 - 	  & 	 - 	  & 	 - 	  \\
 \object{0333$+$321} 	  & 	 NRAO 140 	  & 	 LPQ 	  & 	 17.28 	  & 	 4.9 	  & 	 1.9 	  & 	 0.8 	  & 	 25.37 	  & 	 $4.2^d$ 	  & 	 3.8 	  & 	 $0.7^c$ 	  & 	 N 	  & 	 N 	  & 	 N 	  & 	 N 	  \\
 \object{0336$-$019} 	  & 	 CTA 026 	  & 	 HPQ 	  & 	 15.36 	  & 	 $4.3^w$ 	  & 	 2.5 	  & 	 1.7 	  & 	 16.72 	  & 	 $1.7^w$ 	  & 	 6.3 	  & 	 - 	  & 	 12.28 	  & 	 - 	  & 	 - 	  & 	 - 	  \\
 \object{0355$+$508} 	  & 	 NRAO 150 	  & 	 QSO 	  & 	 22.43 	  & 	 - 	  & 	 - 	  & 	 - 	  & 	 25.41 	  & 	 - 	  & 	 - 	  & 	 - 	  & 	 14.95 	  & 	 - 	  & 	 - 	  & 	 0.3,0.7 	  \\
 \object{0415$+$379} 	  & 	 3C 111 	  & 	 GAL 	  & 	 11.50 	  & 	 4.3 	  & 	 2.0 	  & 	 0.7 	  & 	 12.46 	  & 	 4.9 	  & 	 2.2 	  & 	 - 	  & 	 N 	  & 	 N 	  & 	 N 	  & 	 N 	  \\
 \object{0420$-$014} 	  & 	 OA 129 	  & 	 HPQ 	  & 	 22.32 	  & 	 5.5 	  & 	 4.1 	  & 	 - 	  & 	 21.11 	  & 	 $5.5^c$ 	  & 	 2.7 	  & 	 - 	  & 	 14.93 	  & 	 4.3 	  & 	 2.5 	  & 	 0.7 	  \\
 \object{0422$+$004} 	  & 	 OF 038 	  & 	 BLO 	  & 	 19.25 	  & 	 6.9 	  & 	 2.4 	  & 	 - 	  & 	 19.20 	  & 	 $6.9^c$ 	  & 	 2.7 	  & 	 1.7,0.4 	  & 	 N 	  & 	 N 	  & 	 N 	  & 	 N 	  \\
 \object{0430$+$052} 	  & 	 3C 120 	  & 	 GAL 	  & 	 22.42 	  & 	 4.3 	  & 	 4.4 	  & 	 1.4,0.5 	  & 	 24.12 	  & 	 $4.3^c$ 	  & 	 4.7 	  & 	 1.4,0.3 	  & 	 14.95 	  & 	 2.7 	  & 	 4.0 	  & 	 0.7 	  \\
 \object{0446$+$112} 	  & 	 PKS 0446+112 	  & 	 GAL 	  & 	 15.97 	  & 	 2.2 	  & 	 2.5 	  & 	 - 	  & 	 16.72 	  & 	 2.2 	  & 	 2.0 	  & 	 - 	  & 	 N 	  & 	 N 	  & 	 N 	  & 	 N 	  \\
 \object{0458$-$020} 	  & 	 PKS 0458-020 	  & 	 HPQ 	  & 	 16.16 	  & 	 6.9 	  & 	 1.5 	  & 	 - 	  & 	 17.08 	  & 	 $5.5^c$ 	  & 	 3.1 	  & 	 0.9 	  & 	 N 	  & 	 N 	  & 	 N 	  & 	 N 	  \\
 \object{0528$+$134} 	  & 	 PKS 0528+134 	  & 	 LPQ 	  & 	 15.96 	  & 	 - 	  & 	 - 	  & 	 - 	  & 	 16.92 	  & 	 3.1 	  & 	 3.9 	  & 	 - 	  & 	 13.47 	  & 	 2.7 	  & 	 2.0 	  & 	  	  \\
 \object{0552$+$398} 	  & 	 DA 193 	  & 	 LPQ 	  & 	 14.04 	  & 	 6.9 	  & 	 2.0 	  & 	 - 	  & 	 14.96 	  & 	 8.7 	  & 	 1.7 	  & 	 0.9 	  & 	 13.41 	  & 	 5.5 	  & 	 1.5 	  & 	 0.9 	  \\
 \object{0642$+$449} 	  & 	 OH 471 	  & 	 LPQ 	  & 	 23.75 	  & 	 - 	  & 	 - 	  & 	 0.9,0.4 	  & 	 24.10 	  & 	 - 	  & 	 - 	  & 	 - 	  & 	 N 	  & 	 N 	  & 	 N 	  & 	 N 	  \\
 \object{0716$+$714} 	  & 	  	  & 	 BLO 	  & 	 15.90 	  & 	 5.5 	  & 	 2.1 	  & 	 - 	  & 	 16.74 	  & 	 4.3 	  & 	 1.8 	  & 	 0.5 	  & 	 8.82 	  & 	 - 	  & 	 - 	  & 	 - 	  \\
 \object{0735$+$178} 	  & 	 PKS 0735+17 	  & 	 BLO 	  & 	 23.75 	  & 	 - 	  & 	 - 	  & 	 - 	  & 	 24.04 	  & 	 - 	  & 	 - 	  & 	 - 	  & 	 14.94 	  & 	 - 	  & 	 - 	  & 	 - 	  \\
 \object{0736$+$017} 	  & 	  	  & 	 HPQ 	  & 	 21.22 	  & 	 $2.7^c$ 	  & 	 4.0 	  & 	 0.5 	  & 	 21.90 	  & 	 3.4 	  & 	 4.0 	  & 	 - 	  & 	 11.47 	  & 	 1.4 	  & 	 4.0 	  & 	 - 	  \\
 \object{0754$+$100} 	  & 	 OI 090.4 	  & 	 BLO 	  & 	 20.33 	  & 	 $5.5^c$ 	  & 	 3.1 	  & 	 - 	  & 	 25.37 	  & 	 $5.5^c$ 	  & 	 2.6 	  & 	 0.9 	  & 	 N 	  & 	 N 	  & 	 N 	  & 	 N 	  \\
 \object{0804$+$499} 	  & 	  	  & 	 HPQ 	  & 	 15.97 	  & 	 $5.5^c$ 	  & 	 1.7 	  & 	 $2.2,0.7^c$ 	  & 	 16.74 	  & 	 $5.5^c$ 	  & 	 1.2 	  & 	 $2.2,0.7^c$ 	  & 	 N 	  & 	 N 	  & 	 N 	  & 	 N 	  \\
 \object{0814$+$425} 	  & 	  	  & 	 BLO 	  & 	 16.12 	  & 	 $4.3^{f,w}$	  & 	 2.8 	  & 	 1.1 	  & 	 16.89 	  & 	 - 	  & 	 - 	  & 	 - 	  & 	 N 	  & 	 N 	  & 	 N 	  & 	 N 	  \\
 \object{0827$+$243} 	  & 	 OJ 248 	  & 	 LPQ 	  & 	 10.73 	  & 	 $1.4^{f,w}$ 	  & 	 4.2 	  & 	 - 	  & 	 11.36 	  & 	 $1.4^{f,r}$ 	  & 	 6.4 	  & 	 - 	  & 	 N 	  & 	 N 	  & 	 N 	  & 	 N 	  \\
 \object{0836$+$710} 	  & 	 4C 71.07 	  & 	 LPQ 	  & 	 16.06 	  & 	 - 	  & 	 - 	  & 	 - 	  & 	 16.87 	  & 	 $2.7^{f,w}$ 	  & 	 4.4 	  & 	 - 	  & 	 N 	  & 	 N 	  & 	 N 	  & 	 N 	  \\
 \object{0851$+$202} 	  & 	 OJ 287 	  & 	 BLO 	  & 	 23.27 	  & 	 1.4 	  & 	 10.2 	  & 	 0.4 	  & 	 24.82 	  & 	 1.7 	  & 	 9.7 	  & 	 0.3 	  & 	 14.95 	  & 	 1.4 	  & 	 7.2 	  & 	 - 	  \\
 \object{0906$+$430} 	  & 	 3C 216 	  & 	 HPQ 	  & 	 20.32 	  & 	 - 	  & 	 - 	  & 	 - 	  & 	 21.93 	  & 	 - 	  & 	 - 	  & 	 - 	  & 	 N 	  & 	 N 	  & 	 N 	  & 	 N 	  \\
 \object{0923$+$392} 	  & 	 4C 39.25 	  & 	 LPQ 	  & 	 23.87 	  & 	 - 	  & 	 - 	  & 	 - 	  & 	 24.79 	  & 	 - 	  & 	 - 	  & 	 - 	  & 	 14.91 	  & 	 - 	  & 	 - 	  & 	 - 	  \\
 \object{0945$+$408} 	  & 	 4C 40.24 	  & 	 LPQ 	  & 	 15.82 	  & 	 - 	  & 	 - 	  & 	 - 	  & 	 16.73 	  & 	 - 	  & 	 - 	  & 	 - 	  & 	 N 	  & 	 N 	  & 	 N 	  & 	 N 	  \\
 \object{0953$+$254} 	  & 	  	  & 	 LPQ 	  & 	 15.96 	  & 	 - 	  & 	 - 	  & 	 - 	  & 	 16.71 	  & 	 - 	  & 	 - 	  & 	 - 	  & 	 N 	  & 	 N 	  & 	 N 	  & 	 N 	  \\
 \object{0954$+$556} 	  & 	 S4 0954+556 	  & 	 HPQ 	  & 	 15.82 	  & 	 - 	  & 	 - 	  & 	 - 	  & 	 16.70 	  & 	 - 	  & 	 - 	  & 	 - 	  & 	 N 	  & 	 N 	  & 	 N 	  & 	 N 	  \\
 \object{0954$+$658} 	  & 	 S4 0954+65 	  & 	 BLO 	  & 	 16.14 	  & 	 $5.5^{c,f,w}$ 	  & 	 2.3 	  & 	 1.4 	  & 	 21.90 	  & 	 - 	  & 	 - 	  & 	 0.9 	  & 	 N 	  & 	 N 	  & 	 N 	  & 	 N 	  \\
 \object{1055$+$018} 	  & 	 OL 093 	  & 	 HPQ 	  & 	 22.43 	  & 	 4.3 	  & 	 4.4 	  & 	 - 	  & 	 24.08 	  & 	 $4.3^r$ 	  & 	 4.4 	  & 	 0.7 	  & 	 12.74 	  & 	 2.7 	  & 	 2.6 	  & 	 - 	  \\
 \object{1101$+$384} 	  & 	 Mrk 421 	  & 	 BLO 	  & 	 15.01 	  & 	 - 	  & 	 - 	  & 	 - 	  & 	 19.13 	  & 	 - 	  & 	 - 	  & 	 - 	  & 	 N 	  & 	 N 	  & 	 N 	  & 	 N 	  \\
 \object{1147$+$245} 	  & 	 B2 1147+24 	  & 	 BLO 	  & 	 15.82 	  & 	 - 	  & 	 - 	  & 	 - 	  & 	 15.79 	  & 	 - 	  & 	 - 	  & 	 - 	  & 	 N 	  & 	 N 	  & 	 N 	  & 	 N 	  \\
 \object{1156$+$295} 	  & 	 4C 29.45 	  & 	 HPQ 	  & 	 20.18 	  & 	 3.4 	  & 	 1.7 	  & 	 2.2 	  & 	 21.92 	  & 	 3.4 	  & 	 3.0 	  & 	 1.7 	  & 	 12.52 	  & 	 - 	  & 	 - 	  & 	 - 	  \\
 \object{1219$+$285} 	  & 	 ON 231 	  & 	 BLO 	  & 	 23.89 	  & 	 - 	  & 	 - 	  & 	 - 	  & 	 24.11 	  & 	 8.7 	  & 	 2.5 	  & 	 - 	  & 	 N 	  & 	 N 	  & 	 N 	  & 	 N 	  \\
 \object{1222$+$216} 	  & 	 PKS 1222+216 	  & 	 QSO 	  & 	 10.55 	  & 	 2.2 	  & 	 2.4 	  & 	 - 	  & 	 11.32 	  & 	 2.2 	  & 	 2.3 	  & 	 - 	  & 	 N 	  & 	 N 	  & 	 N 	  & 	 N 	  \\
 \object{1226$+$023} 	  & 	 3C 273 	  & 	 LPQ 	  & 	 24.03 	  & 	 $6.9^r$ 	  & 	 3.2 	  & 	 2.2 	  & 	 25.30 	  & 	 $7.7^{c,r}$ 	  & 	 2.7 	  & 	 - 	  & 	 14.96 	  & 	 $4.3^{c,d}$ 	  & 	 3.0 	  & 	 - 	  \\
 \object{1253$-$055} 	  & 	 3C 279 	  & 	 HPQ 	  & 	 24.01 	  & 	 - 	  & 	 - 	  & 	 - 	  & 	 25.30 	  & 	 - 	  & 	 - 	  & 	 - 	  & 	 14.93 	  & 	 - 	  & 	 - 	  & 	 - 	  \\
 \object{1308$+$326} 	  & 	 AU CV n 	  & 	 BLO 	  & 	 22.28 	  & 	 10.9 	  & 	 2.0 	  & 	 - 	  & 	 24.00 	  & 	 9.7 	  & 	 2.3 	  & 	 - 	  & 	 10.04 	  & 	 $4.3^c$ 	  & 	 2.3 	  & 	 - 	  \\
 \object{1413$+$135} 	  & 	  	  & 	 BLO 	  & 	 15.36 	  & 	 3.1 	  & 	 4.5 	  & 	 - 	  & 	 16.13 	  & 	 3.1 	  & 	 2.7 	  & 	 - 	  & 	 10.12 	  & 	 $2.7^c$ 	  & 	 1.8 	  & 	 - 	  \\
 \object{1418$+$546} 	  & 	 OQ 530 	  & 	 BLO 	  & 	 21.05 	  & 	 - 	  & 	 - 	  & 	 - 	  & 	 21.96 	  & 	 - 	  & 	 - 	  & 	 - 	  & 	 N 	  & 	 N 	  & 	 N 	  & 	 N 	  \\
 \object{1502$+$106} 	  & 	 OR 103 	  & 	 HPQ 	  & 	 16.11 	  & 	 $6.9^w$ 	  & 	 2.0 	  & 	 - 	  & 	 23.47 	  & 	 $6.9^w$ 	  & 	 1.7 	  & 	 1.7 	  & 	 N 	  & 	 N 	  & 	 N 	  & 	 N 	  \\
 \object{1510$-$089} 	  & 	 PKS 1510-089 	  & 	 HPQ 	  & 	 19.94 	  & 	 - 	  & 	 - 	  & 	 - 	  & 	 21.93 	  & 	 - 	  & 	 - 	  & 	 - 	  & 	 14.05 	  & 	 $3.4^c$ 	  & 	 2.4 	  & 	 - 	  \\
 \object{1538$+$149} 	  & 	 4C 14.60 	  & 	 BLO 	  & 	 20.33 	  & 	 $2.7^{c,f,w}$ 	  & 	 4.0 	  & 	 - 	  & 	 21.93 	  & 	 - 	  & 	 - 	  & 	 - 	  & 	 N 	  & 	 N 	  & 	 N 	  & 	 N 	  \\
 \object{1606$+$106} 	  & 	 4C 10.45 	  & 	 LPQ 	  & 	 11.28 	  & 	 5.5 	  & 	 1.5 	  & 	 - 	  & 	 12.10 	  & 	 4.9 	  & 	 1.9 	  & 	 - 	  & 	 N 	  & 	 N 	  & 	 N 	  & 	 N 	  \\
 \object{1611$+$343} 	  & 	 DA 406 	  & 	 LPQ 	  & 	 15.96 	  & 	 4.3,10.9 	  & 	 3.1 	  & 	 - 	  & 	 16.83 	  & 	 4.3,10.9 	  & 	 3.1 	  & 	 - 	  & 	 N 	  & 	 N 	  & 	 N 	  & 	 N 	  \\
 \object{1633$+$382} 	  & 	 4C 38.41 	  & 	 LPQ 	  & 	 22.43 	  & 	 $3.4^c$ 	  & 	 4.8 	  & 	 - 	  & 	 24.08 	  & 	 $3.4^c$ 	  & 	 5.1 	  & 	 - 	  & 	 N 	  & 	 N 	  & 	 N 	  & 	 N 	  \\
 \object{1637$+$574} 	  & 	 OS 562 	  & 	 LPQ 	  & 	 20.17 	  & 	 $3.4^r$ 	  & 	 4.0 	  & 	 - 	  & 	 21.94 	  & 	 $3.9^r$ 	  & 	 2.8 	  & 	 - 	  & 	 N 	  & 	 N 	  & 	 N 	  & 	 N 	  \\
 \object{1641$+$399} 	  & 	 3C 345 	  & 	 HPQ 	  & 	 24.00 	  & 	 $8.7^r$ 	  & 	 2.5 	  & 	 - 	  & 	 24.80 	  & 	 $8.7^r$ 	  & 	 2.7 	  & 	 - 	  & 	 14.93 	  & 	 - 	  & 	 - 	  & 	 - 	  \\
 \object{1652$+$398} 	  & 	 Mrk 501 	  & 	 BLO 	  & 	 16.11 	  & 	 - 	  & 	 - 	  & 	 - 	  & 	 16.91 	  & 	 - 	  & 	 - 	  & 	 - 	  & 	 N 	  & 	 N 	  & 	 N 	  & 	 N 	  \\
 \object{1725$+$044} 	  & 	 PKS 1725+044 	  & 	 QSO 	  & 	 12.70 	  & 	 - 	  & 	 - 	  & 	 - 	  & 	 13.62 	  & 	 - 	  & 	 - 	  & 	 - 	  & 	 N 	  & 	 N 	  & 	 N 	  & 	 N 	  \\
 \object{1739$+$522} 	  & 	 S4 1739+52 	  & 	 HPQ 	  & 	 15.84 	  & 	 5.5 	  & 	 2.4 	  & 	 1.1 	  & 	 16.87 	  & 	 $4.3^{c,d}$ 	  & 	 2.5 	  & 	 1.1 	  & 	 N 	  & 	 N 	  & 	 N 	  & 	 N 	  \\
 \object{1741$-$038} 	  & 	 PKS 1741-038 	  & 	 HPQ 	  & 	 16.03 	  & 	 $5.5^r$ 	  & 	 2.8 	  & 	 - 	  & 	 16.99 	  & 	 $5.5^r$ 	  & 	 2.9 	  & 	 0.5 	  & 	 13.35 	  & 	 6.1 	  & 	 2.0 	  & 	 - 	  \\
 \object{1749$+$096} 	  & 	 PKS 1749+096 	  & 	 BLO 	  & 	 19.86 	  & 	 $2.7^{c,d}$ 	  & 	 4.4 	  & 	 0.4 	  & 	 24.53 	  & 	 $2.7^{c,d}$ 	  & 	 5.0 	  & 	 0.5 	  & 	 14.10 	  & 	 $2.7^{c,d}$ 	  & 	 3.4 	  & 	 - 	  \\
 \object{1803$+$784} 	  & 	 S5 1803+784 	  & 	 BLO 	  & 	 15.91 	  & 	 $3.4^{f,w}$ 	  & 	 2.9 	  & 	 - 	  & 	 16.75 	  & 	 $3.4^{f,w}$ 	  & 	 2.9 	  & 	 - 	  & 	 9.10 	  & 	 - 	  & 	 - 	  & 	 - 	  \\
 \object{1807$+$698} 	  & 	 3C 371.0 	  & 	 BLO 	  & 	 20.17 	  & 	 - 	  & 	 - 	  & 	 - 	  & 	 21.95 	  & 	 - 	  & 	 - 	  & 	 - 	  & 	 N 	  & 	 N 	  & 	 N 	  & 	 N 	  \\
 \object{1823$+$568} 	  & 	 4C 56.27 	  & 	 BLO 	  & 	 15.56 	  & 	 - 	  & 	 - 	  & 	 - 	  & 	 16.70 	  & 	 - 	  & 	 - 	  & 	 - 	  & 	 9.04 	  & 	 - 	  & 	 - 	  & 	 - 	  \\
 \object{1928$+$738} 	  & 	 4C 73.18 	  & 	 LPQ 	  & 	 16.06 	  & 	 $5.5^w$ 	  & 	 1.9 	  & 	 - 	  & 	 16.85 	  & 	 - 	  & 	 - 	  & 	 $0.6^f$ 	  & 	 N 	  & 	 N 	  & 	 N 	  & 	 N 	  \\
 \object{2005$+$403} 	  & 	  	  & 	 QSO 	  & 	 22.29 	  & 	 - 	  & 	 - 	  & 	 - 	  & 	 24.03 	  & 	 - 	  & 	 - 	  & 	 - 	  & 	 N 	  & 	 N 	  & 	 N 	  & 	 N 	  \\
 \object{2007$+$776} 	  & 	 S5 2007+77 	  & 	 BLO 	  & 	 12.38 	  & 	 $3.1^w$ 	  & 	 2.7 	  & 	 - 	  & 	 16.83 	  & 	 - 	  & 	 - 	  & 	 - 	  & 	 N 	  & 	 N 	  & 	 N 	  & 	 N 	  \\
 \object{2021$+$614} 	  & 	 OW 637 	  & 	 LPQ 	  & 	 16.80 	  & 	 - 	  & 	 - 	  & 	 - 	  & 	 21.93 	  & 	 - 	  & 	 - 	  & 	 - 	  & 	 N 	  & 	 N 	  & 	 N 	  & 	 N 	  \\
 \object{2134$+$004} 	  & 	 OX 057 	  & 	 LPQ 	  & 	 22.27 	  & 	 6.9 	  & 	 3.2 	  & 	 - 	  & 	 25.37 	  & 	 - 	  & 	 - 	  & 	 - 	  & 	 N 	  & 	 N 	  & 	 N 	  & 	 N 	  \\
 \object{2136$+$141} 	  & 	  	  & 	 LPQ 	  & 	 15.51 	  & 	 8.7 	  & 	 1.5 	  & 	 - 	  & 	 18.20 	  & 	 6.9 	  & 	 1.4 	  & 	 - 	  & 	 N 	  & 	 N 	  & 	 N 	  & 	 N 	  \\
 \object{2145$+$067} 	  & 	  	  & 	 LPQ 	  & 	 18.37 	  & 	 6.1 	  & 	 2.7 	  & 	 - 	  & 	 19.16 	  & 	 5.5 	  & 	 3.0 	  & 	 - 	  & 	 14.93 	  & 	 - 	  & 	 - 	  & 	 - 	  \\
 \object{2200$+$420} 	  & 	 BL Lac 	  & 	 BLO 	  & 	 24.01 	  & 	 $8.7^c$ 	  & 	 2.5 	  & 	 - 	  & 	 25.44 	  & 	 $7.7^c$ 	  & 	 3.1 	  & 	 - 	  & 	 14.96 	  & 	 - 	  & 	 - 	  & 	 0.9 	  \\
 \object{2201$+$315} 	  & 	 4C 31.63 	  & 	 LPQ 	  & 	 19.25 	  & 	 4.3 	  & 	 2.5 	  & 	 - 	  & 	 22.54 	  & 	 4.3 	  & 	 4.4 	  & 	 0.5 	  & 	 N 	  & 	 N 	  & 	 N 	  & 	 N 	  \\
 \object{2223$-$052} 	  & 	 3C 446 	  & 	 BLO 	  & 	 19.35 	  & 	 9.7 	  & 	 1.9 	  & 	 - 	  & 	 19.22 	  & 	 9.7 	  & 	 1.8 	  & 	 - 	  & 	 14.94 	  & 	 8.7 	  & 	 1.5 	  & 	 - 	  \\
 \object{2230$+$114} 	  & 	 CTA 102 	  & 	 HPQ 	  & 	 19.25 	  & 	 8.7 	  & 	 1.9 	  & 	 0.8 	  & 	 19.21 	  & 	 6.9 	  & 	 2.7 	  & 	 0.9 	  & 	 14.50 	  & 	 7.7 	  & 	 1.6 	  & 	 - 	  \\
 \object{2234$+$282} 	  & 	  	  & 	 HPQ 	  & 	 15.97 	  & 	 - 	  & 	 - 	  & 	 - 	  & 	 15.83 	  & 	 - 	  & 	 - 	  & 	 - 	  & 	 N 	  & 	 N 	  & 	 N 	  & 	 N 	  \\
 \object{2251$+$158} 	  & 	 3C 454.3 	  & 	 HPQ 	  & 	 24.01 	  & 	 $6.1^c$ 	  & 	 3.6 	  & 	 - 	  & 	 24.48 	  & 	 $6.1^c$ 	  & 	 3.6 	  & 	 1.4 	  & 	 14.95 	  & 	 - 	  & 	 - 	  & 	 - 	  \\
\hline 
\end{longtable}
\begin{list}{}{\setlength{\leftmargin}{45pt}}
\item \footnotesize{$^c$ = complex structure in the wavelet plot.}
\item \footnotesize{$^w$ = weak timescale.}
\item \footnotesize{$^f$ = faint source.}
\item \footnotesize{$^r$ = rising trend in the timescale.}
\item \footnotesize{$^d$ = declining trend in the timescale.}
\item \footnotesize{- = timescale not determined.}
\item \footnotesize{N = not enough data for wavelet analysis.}
\end{list}
}
\end{landscape}
}
\end{document}